\documentclass{article}
\usepackage{spconf}
\usepackage{times}          
\usepackage[T1]{fontenc}
\usepackage[utf8]{inputenc}
\usepackage{amsmath,amssymb}
\usepackage{graphicx}
\usepackage{booktabs}
\usepackage{url}
\usepackage{cite}

\graphicspath{{figures/}}

\newcommand{\Sbb}{\hat{S}^{\mathrm{bb}}}
\newcommand{\Sdf}{\hat{S}^{\mathrm{df}}}
\newcommand{\GMACs}{\mathrm{GMAC\,s^{-1}}}

\title{REAL-TIME MUSIC SOURCE SEPARATION ON A LOW-POWER AUDIO DSP}

\name{Jianan Li, Li Liu, Ken Malsky and Gabby Yi}
\address{Analog Devices, Inc., Norwood, MA, USA\\
\texttt{\{jianan.li,\,li.liu,\,ken.malsky,\,gabby.yi\}@analog.com}}

\begin{document}
\ninept
\maketitle

\begin{abstract}
Real-time music source separation is validated on desktop CPUs and GPUs. Does any
published system fit the embedded audio hardware it targets? On a commercial audio DSP
($2$\,MB SRAM, $2.07$\,$\GMACs$ measured), none does, and the constraints eliminate
different models: memory rules out the $16$--$51$\,M parameter TasNet/X-UMX family,
per-frame compute rules out RT-STT, needing $5.5\times$ the available MAC rate. Parameter
count predicts neither: weight reuse spans $1\times$ to $345\times$. We then build one
that fits. Training on \emph{continuous} rather than block-padded convolution context
proves essential: a model scoring $3.93$\,dB block-wise otherwise collapses to silence
within $2$\,s frame-by-frame. A gated complex FIR \emph{deep~filter} adds a latency knob,
gaining $0.38$\,dB even when strictly causal. It reaches $4.70$\,dB cSDR on MUSDB18-HQ and
runs in $10.43$\,ms of an $11.6$\,ms hop, $0.5$--$0.7$\,dB behind systems that do not fit.
\end{abstract}


\section{Introduction}
\label{sec:intro}
Deep learning has transformed music source separation (MSS)~\cite{mdx21,sdx23}, but
the strongest systems are large and offline, consuming a whole track at
once~\cite{htdemucs,bsrnn,bsrope}. A smaller literature targets the \emph{real-time}
regime: HS-TasNet~\cite{hstasnet} demixes at $23$\,ms latency and
RT-STT~\cite{rtstt} matches that latency with ${\sim}0.4$\,M parameters, alongside
work on accompaniment separation~\cite{mmdensenet} and singing-voice
cancellation~\cite{stereosvc}. Both leading systems are explicit that latency is not
the only constraint: each separates \emph{algorithmic latency} from
\emph{computational efficiency} and reports per-frame processing time, and HS-TasNet
notes that Conv-TasNet~\cite{convtasnet} is slow in practice despite having ``only
9\,M parameters''.

Those times are measured on an i7-class CPU and an RTX-class
GPU~\cite{hstasnet,rtstt}. A separator built for hearing aids, in-ear monitors or a
live-sound processor runs instead on a low-power audio DSP: a fixed
multiply--accumulate (MAC) rate, a hard per-frame deadline, and a few megabytes of
on-chip SRAM. ``$3.9$\,ms on an RTX 3080Ti'' does not transfer. Efficiency-oriented
separation has reached mobile GPUs~\cite{mobilesep} and speech enhancement runs
routinely on embedded parts~\cite{deepfilternet}, but MSS is harder at the same
budget (four correlated outputs, and a decoder that reconstructs every stem), and we
know of no MSS system evaluated against, or demonstrated on, such a device.

We close that gap. Section~\ref{sec:budget} states deployability as two
hardware-independent constraints, weight memory and MAC per frame, and evaluates the
published systems against a commercial budget;
Sections~\ref{sec:method}--\ref{sec:results} design, train and deploy a separator that
fits. Along the way we report a streaming failure of independent interest: block-padded
convolution training makes continuous inference out-of-distribution, and a model
scoring $3.93$\,dB block-wise collapses to silence within ${\sim}2$\,s frame-by-frame.

The backbone follows the TFC-TDF U-Net lineage~\cite{tfctdf,choithesis} as
specialised for real time by RT-STT~\cite{rtstt}, itself derived from
DTTNet~\cite{dttnet}; band-split models~\cite{bsrnn,bsrope,moiseslight} and
low-latency speech separation~\cite{lowlatmc} are the other relevant strands, and
efficiency-driven designs such as SCNet~\cite{scnet} reduce cost at desktop scale
rather than to an embedded budget. On top we
add \emph{deep filtering}~\cite{deepfilter,deepfilternet}, which replaces the
point-wise complex mask with a short complex FIR filter per time--frequency bin
convolved along time; used offline as a post-filter on Hybrid Demucs~\cite{cadenzadf}
and for impulsive/stationary separation~\cite{is3}, we use it causally, gate it per
source so the network can decline it, and treat its past/future tap split as an
explicit latency control.

\section{The Embedded Budget}
\label{sec:budget}

\noindent\textbf{Target.} We use the Analog Devices SHARC-FX (ADSP-21835) as a
representative low-power audio DSP: $1$\,GHz, $512$\,kB L1, $2$\,MB on-chip L2. Its
published peak is $24$\,GFLOPS, or $8$\,$\GMACs$ in $32$-bit float and $16$\,$\GMACs$ in
$16$-bit fixed point. Our hand-scheduled floating-point runtime sustains
$2.07$\,$\GMACs$ on this part (measured, \S\ref{sec:deploy}). At $44.1$\,kHz with a
$512$-sample hop, the $23$\,ms operating point of both baselines, this gives $86.1$
frames\,s$^{-1}$ and an $11.6$\,ms per-frame deadline. The part has a DDR interface, so
weights could live off-chip; we exclude that deliberately, because they are re-read every
frame, so streaming them puts a recurring transfer on the critical path of that deadline:
our $501$\,kB of weights at $86.1$ frames\,s$^{-1}$ is ${\sim}44$\,MB\,s$^{-1}$ sustained
for as long as the device runs. Double-buffered DMA hides the latency but not the
bandwidth, and external
memory adds board cost and power to a product that chose this class of part
to avoid both. $M_{\mathrm{L2}}$ is thus a design constraint, not a hardware limit, and
the analysis is parameterised by $(M_{\mathrm{L2}}, R)$ so it can be re-run for any
target.

\noindent\textbf{Peak rate is not deployable rate.} A datasheet peak assumes every
issue slot retires a useful MAC: no cache miss, no carried state, no serialisation.
Budgeting against it would be a category error. Our $2.07$\,$\GMACs$ is $26\%$ of the
float peak and is hand-scheduled; a TFLite-Micro port on the same part spent
${\sim}90\%$ of its time moving data. We therefore compare
\emph{required} against \emph{measured achievable} rate. Granting a baseline $16$-bit
fixed point at our efficiency gives it $4.2$\,$\GMACs$; RT-STT still needs $2.7\times$
that, so the verdict does not rest on denying baselines fixed-point arithmetic.

\noindent\textbf{Two constraints.} A model is deployable only if it satisfies both:
\begin{align}
\text{memory:} \quad & W \cdot b \;\le\; M_{\mathrm{L2}}, \\
\text{compute:} \quad & C_{\mathrm{frame}} \cdot f_{\mathrm{s}}/H \;\le\; R,
\end{align}
where $W$ is the parameter count, $b$ bytes per weight, $C_{\mathrm{frame}}$ the
MAC per frame, $H$ the hop and $R$ the sustained MAC rate. We evaluate the memory
constraint from published parameter counts under the \emph{most generous} assumption
available to those systems, $b=1$ (int8), which none of them claims; activations and
streaming state are excluded, so the true gap is larger.

\noindent\textbf{Weight reuse, and why $W$ does not predict $C_{\mathrm{frame}}$.}
For a fully-connected or recurrent layer evaluated once per frame, each weight takes
part in exactly one multiply--accumulate, so $C_{\mathrm{frame}}\!=\!W$ identically.
Convolution over frequency breaks this: a kernel is reused at every bin. Define the
reuse factor $\rho = C_{\mathrm{frame}}/W$. It is $\rho\!\approx\!1$ for the
TasNet/X-UMX family, which is dominated by frame-rate LSTMs and dense layers, and
$\rho\!=\!345$ for RT-STT, whose $383$\,k weights are evaluated across $384$
frequency bins every frame. Because $\rho$ spans two and a half orders of magnitude
across Table~\ref{tab:feas}, no ordering by parameter count can predict per-frame
cost. This also gives the baselines' compute for free: at $\rho\!\approx\!1$ the
published $W$ \emph{is} a per-frame MAC estimate, and a lower bound once their
convolutional front ends are added. We validate the identity on X-UMX, whose
architecture is fully specified~\cite{hstasnet}: summing its layers gives
$31.5$\,M MAC/frame against $31$\,M reported parameters.

\begin{figure}[t]
\centering
\includegraphics[width=\columnwidth]{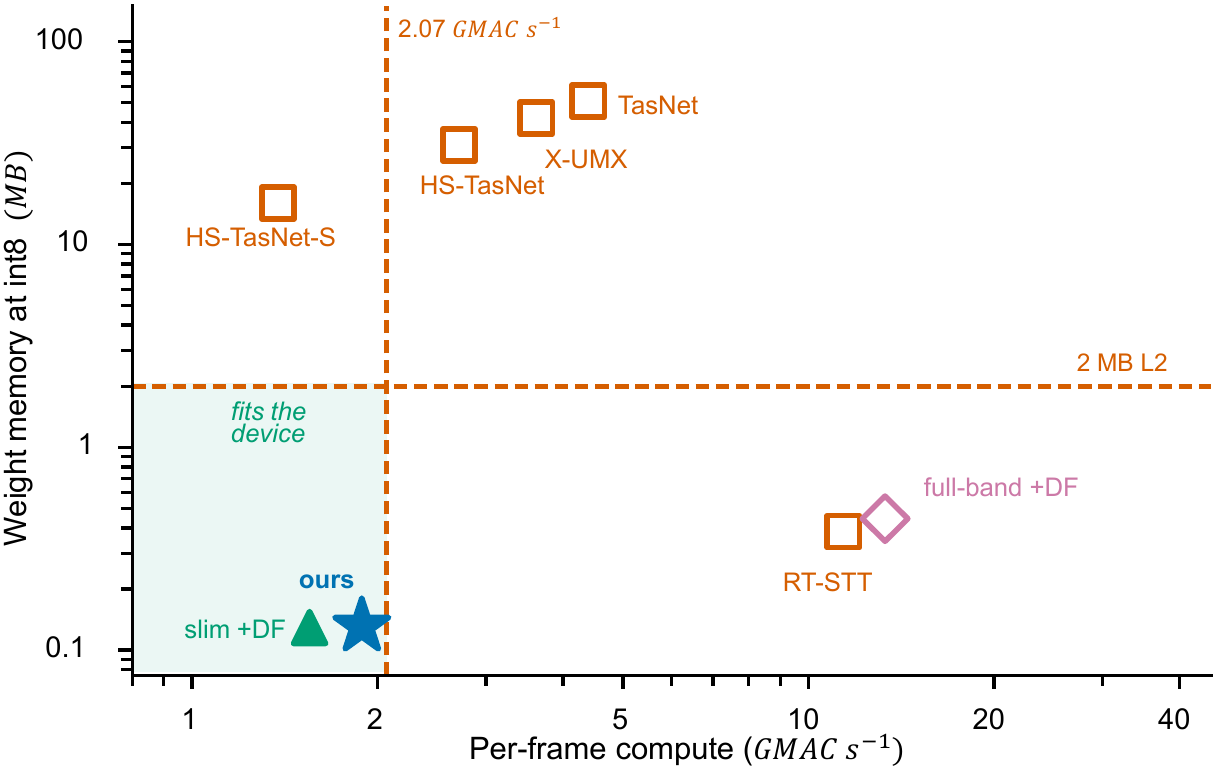}
\caption{Published real-time MSS against a commercial audio DSP: both budgets as
axes, so the deployable set is the shaded box. The two constraints eliminate
different systems (HS-TasNet-S meets the frame deadline but overruns L2, RT-STT the
reverse), and parameter count predicts neither. Quality is given in
Table~\ref{tab:feas}.}
\label{fig:feas}
\end{figure}

\noindent\textbf{Result.} Table~\ref{tab:feas} and Fig.~\ref{fig:feas} give the
verdict: no published system satisfies both constraints, and the two constraints
eliminate different models. Weight memory rules out X-UMX, TasNet and both HS-TasNet
variants by $8.0$--$25.5\times$, a gap no scheduling or quantisation strategy closes.
Compute independently rules out RT-STT, which passes memory easily at $19\%$ of L2
and then needs $11.4$\,$\GMACs$, $5.5\times$ what the device sustains. Three of the
four large models also exceed the compute budget, by $1.3$--$2.1\times$: at
$\rho\!\approx\!1$ being large \emph{is} being expensive per frame.

HS-TasNet-S is the instructive exception: at $16$\,M parameters it is the only prior
system that would meet the frame deadline ($67\%$ of budget), yet it overruns L2 by
$8.0\times$. RT-STT is its mirror image. The constraints are close to anti-correlated
across architecture families, which is why no single size proxy can summarise
deployability: RT-STT carries $2.9\times$ the parameters of our deployed model and
$6\times$ its per-frame cost, at the same latency.

\noindent\textbf{Scope.} The verdict is scoped to audio DSPs of this class, where this
part sits at the favourable end: $14.43$\,AudioMark/MHz against $3.63$ for a Cortex-M55
and $0.76$ for a Cortex-M4~\cite{audiomark}, so it only hardens on M-class targets. It
would not hold on an application processor with a vector backend.

\begin{table}[t]
\caption{Published real-time MSS against a commercial audio DSP budget
($2$\,MB L2, $2.07$\,$\GMACs$ measured, $86.1$ frames\,s$^{-1}$); memory assumes int8,
the most generous case for each system, while our weights are float32. Quality and parameters
are from~\cite{hstasnet} (rows 1--4) and~\cite{rtstt} (RT-STT).
$^{*}$computed from the published architecture; $^{\dagger}$from the
$\rho\!\approx\!1$ identity, hence a lower bound; remaining rows profiled directly.}
\label{tab:feas}
\centering
\setlength{\tabcolsep}{3.2pt}
\begin{tabular}{lccccc}
\toprule
Model & cSDR & Params & Mem. & GMAC/s & $\rho$ \\
\midrule
X-UMX          & 3.93 & 31\,M  & $15.5\times$ & $2.71^{*}$ & 1 \\
TasNet         & 4.40 & 51\,M  & $25.5\times$ & $4.39^{\dagger}$ & 1 \\
HS-TasNet-S    & 4.48 & 16\,M  & $8.0\times$  & $1.38^{\dagger}$ & 1 \\
HS-TasNet      & 4.65 & 42\,M  & $21.0\times$ & $3.62^{\dagger}$ & 1 \\
RT-STT         & 5.17 & 383\,K & ok           & 11.39 & 345 \\
\midrule
full-band $+$DF (ours) & 5.49 & 444\,K & ok & 13.31 & 348 \\
slim $+$DF (ours)      & 4.34 & 129\,K & ok & \textbf{1.55} & 140 \\
deployed (ours)        & 4.70 & 131\,K & ok & \textbf{1.89} & 167 \\
\bottomrule
\end{tabular}
\end{table}

\section{Method}
\label{sec:method}

\begin{figure*}[t]
\centering
\includegraphics[width=\textwidth]{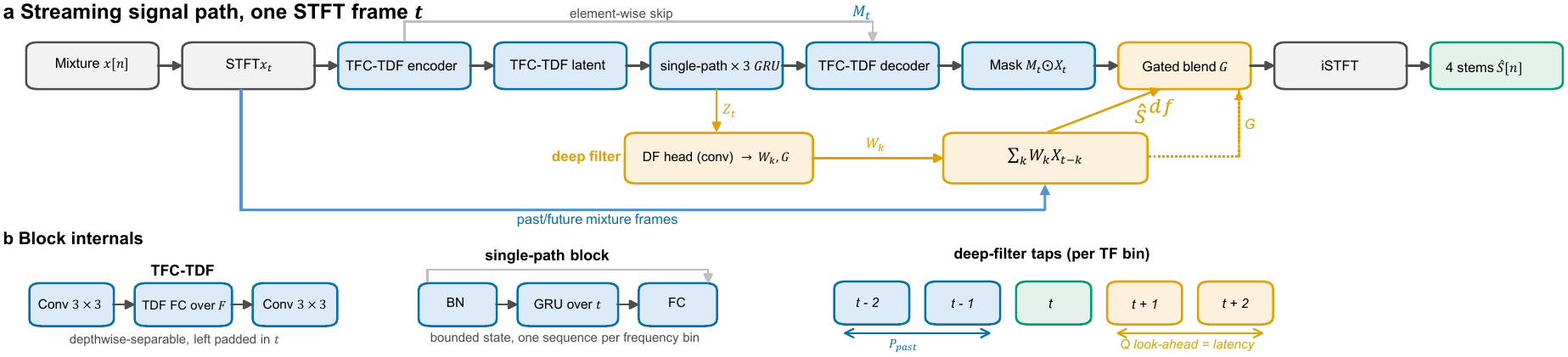}
\caption{(\textbf{a}) Streaming signal path for one STFT frame: the causal TFC-TDF
U-Net emits a complex ratio mask $M_t$, a deep-filter head predicts per-bin complex
FIR taps $W_k$ and a gate $G$ from the latent $Z_t$, and the two estimates are
blended before the iSTFT. (\textbf{b}) Block internals; the look-ahead order $Q$ is
the only source of added algorithmic latency.}
\label{fig:arch}
\end{figure*}

Fig.~\ref{fig:arch} shows the system; panel (a) gives the layer order, and all
convolutions are depthwise-separable and left-padded in time. We describe only what
the budget forced.

\subsection{Backbone}
The separator is a causal TFC-TDF U-Net~\cite{tfctdf,rtstt} operating on the
real-valued STFT of the mixture ($n_{\mathrm{fft}}{=}1024$, hop $H{=}512$, no centre
padding, so frame $t$ depends only on past samples). The $1024$-point transform gives
$513$ bins; following~\cite{rtstt} we keep the lowest $F{=}384$ ($0$--$16.5$\,kHz) and
reconstruct $16.5$--$22.05$\,kHz by scaling the mixture's high band by a per-source gain
from the top in-band bins. The crop is a compute dial, MAC scaling linearly in $F$; we
also report an $F{=}192$ ($0$--$8.27$\,kHz) variant.

Two choices are dictated by streaming rather than accuracy. \emph{(i)}~The recurrence
is a GRU, whose state $h_t=(1-z_t)h_{t-1}+z_t\tilde{h}_t$ is a convex combination of
bounded terms and so stays in $[-1,1]$ by construction, whereas an LSTM cell state
$c_t=f_t c_{t-1}+i_t g_t$ is an unbounded running sum. Streaming two trained
checkpoints for $120$\,s and tracking $\max|\cdot|$ of the state, the LSTM grows from
$399$ to $962$, still rising at $+2.4$\,s$^{-1}$ with no plateau, while the GRU sits
at exactly $1.00$. An unbounded state eventually saturates the downstream
activations, so an LSTM needs periodic resets, which reintroduce buffering latency. We
did not test whether forget-bias initialisation or recurrent normalisation would bound it
instead; the GRU removes the failure mode by construction, which on a part with no reset
path is the cheaper argument.
\emph{(ii)}~The network emits a complex ratio mask on the input spectrum,
$\Sbb_t = M_t \odot X_t$, so silence maps to exactly silence and the idle noise floor
is zero by construction rather than by training.

\subsection{Deep filter}
From the latent features $Z_t$ a small convolutional head predicts, per source and TF
bin, a complex FIR filter $W_k$ of order $N{=}P{+}Q{+}1$ and a gate $G\in(0,1)$. The head
is two $3\times3$ convolutions (width $32$, BN, ReLU) and a $1\times1$ projection to
$S(2N{+}1)$ channels, so taps are shared across frequency by construction; it adds
$58.5$\,k parameters, $15.2\%$ over the backbone. Real and imaginary parts are each
$\tanh$-bounded, the only stability constraint we impose. At $Q{=}0$ the head's own time
padding is left-only, so the coefficient \emph{predictor} is causal too; without that the
filter is causal but its taps are not. The
filter is applied to the \emph{mixture} spectrum,
\begin{equation}
\Sdf_t \;=\; \sum_{k=-Q}^{P} W_k \odot X_{t-k},
\end{equation}
and blended with the mask estimate,
\begin{equation}
\hat{S}_t \;=\; G \odot \Sbb_t + (1-G) \odot \Sdf_t .
\end{equation}
Because the filter reads the mixture, it too maps silence to silence. The split
between $P$ past and $Q$ future taps sets the added algorithmic latency exactly:
$Q\cdot H/f_{\mathrm{s}} = 11.6Q$\,ms. We use $N{=}5$ with $Q\in\{0,1,2\}$.

\subsection{Training on continuous context}
\label{sec:contconv}
Causal convolutions are normally trained on independent chunks, each zero-padded on
the left; at inference the same layers see a running cache of real past frames.
Stacked layers compound the mismatch: a $3\times3$ kernel depends on two past frames,
so an $L$-layer stack makes the first $2L$ frames of a chunk depend on padding. With
$L\approx8$ layers across encoder, latent and decoder, ${\sim}16$ of the $65$ frames
in a $0.755$\,s chunk, a quarter of them, sit in a regime that never occurs in deployment.

\begin{figure*}[t]
\centering
\includegraphics[width=\textwidth]{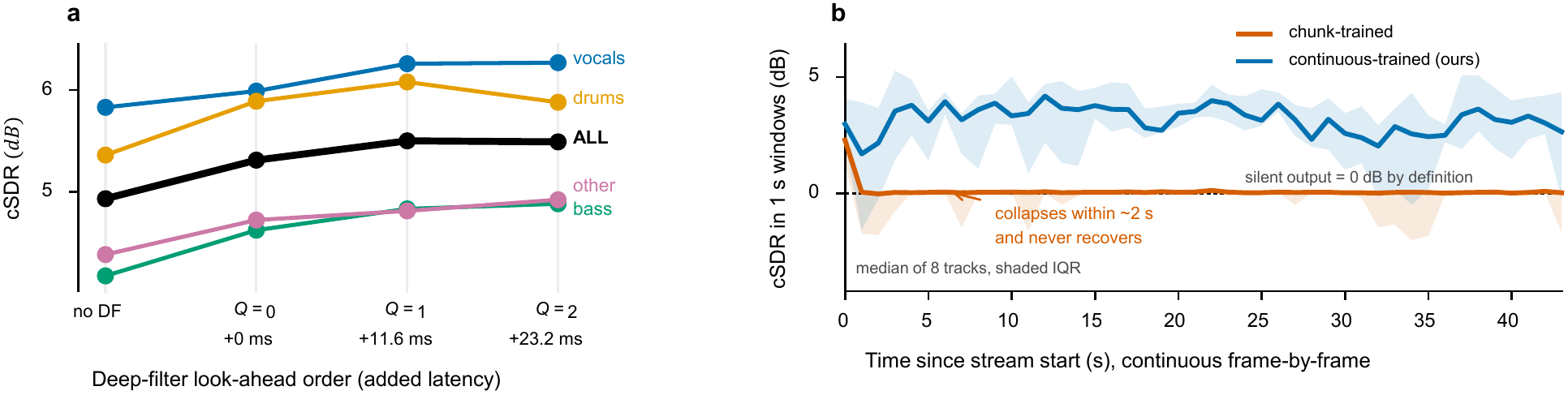}
\caption{(\textbf{a}) Per-stem cSDR across the look-ahead sweep: every stem rises
already at the strictly causal $Q{=}0$, so the gain is complex FIR filtering rather
than future information. (\textbf{b}) Streaming trajectory under continuous
frame-by-frame inference; median of $8$ test tracks, shaded interquartile range. Both
start identically at $t{=}0$, where the convolution cache genuinely is zero, but the
chunk-trained model collapses to silence within ${\sim}2$\,s.}
\label{fig:tradeoff}
\end{figure*}

The consequence is easy to miss. A chunk-trained model scores $3.93$\,dB under the
standard block-wise protocol, which resets state at every block, while the
\emph{same weights} run frame-by-frame collapse to silent output within ${\sim}2$\,s
(Fig.~\ref{fig:tradeoff}b). An ablation isolates the cause: threading recurrent state
across blocks while keeping the per-block zero padding costs only ${\sim}0.2$\,dB, so
it is the convolution context, not the recurrent state. Training on single continuous
segments ($5.8$\,s, one forward pass, no internal padding) restores frame-by-frame
performance to within $0.1$\,dB of block-wise. The recipe is otherwise unremarkable, and
that is the point: one forward pass over a $500$-frame segment cut at a random position,
sources re-mixed across tracks as usual, batch $6$ ($8.8$\,GB on one T4), AdamW at
$10^{-4}$, gradient-norm clip $3.0$, mixed precision; the DF head is warm-started onto a
converged backbone. A truncated-BPTT variant must also overlap consecutive chunks by
$n_{\mathrm{fft}}{-}H$ samples, or the frame grid skips a position at every boundary.

\begin{table}[t]
\caption{Deployable operating points. Time/frame at $1$\,GHz against the $11.6$\,ms
deadline, end to end (signal processing included). Times are measured on the part; the
two slim $+$DF rows are projected from the measured $F{=}192$ rate, network term scaled
by MAC and the $0.28$\,ms of signal processing held fixed.}
\label{tab:deploy}
\centering
\setlength{\tabcolsep}{3.6pt}
\begin{tabular}{lccccc}
\toprule
Model & cSDR & uSDR & MAC/fr & Time/fr & Budget \\
\midrule
slim, no DF      & 4.04 & 4.24 & 15.1\,M & 7.56\,ms & 65\% \\
slim $+$DF (LA0) & 4.10 & 4.30 & 18.0\,M & 8.98\,ms & 77\% \\
slim $+$DF (LA2) & 4.34 & 4.52 & 18.0\,M & 8.98\,ms & 77\% \\
full $+$DF (LA2) & \textbf{4.70} & \textbf{4.70} & 21.9\,M & \textbf{10.43\,ms} & \textbf{90\%} \\
\bottomrule
\end{tabular}
\end{table}

\section{Experimental Setup}
\label{sec:setup}
\noindent\textbf{Data and metrics.} MUSDB18-HQ~\cite{musdb18hq}, evaluated on the full
$50$-song test split at $44.1$\,kHz. We report cSDR (median over $1$\,s windows, then
over songs, silent reference windows gated) and uSDR (whole-song SDR, mean over
songs)~\cite{mdx21,sdx23}. cSDR is a plain energy ratio, not the BSSEval-v4 projection
of \texttt{museval}~\cite{museval,bsseval}: on our $50$ tracks, evaluated on the stereo
pair, they differ by $-1.33$ to $+1.21$\,dB per stem and the sign is not consistent
across stems, so cross-paper comparisons are indicative. To make ours directly
comparable we report both for the deployed model: $4.70$\,dB cSDR is $4.48$\,dB under
BSSEval-v4, per stem $5.09/5.65/3.87/4.18$ against $6.30/5.21/3.56/2.85$ for
vocals/drums/bass/other. ``ALL'' is the stem mean.

\noindent\textbf{Training.} AdamW at lr $10^{-4}$ with an $L_1$ waveform loss and the
augmentation of~\cite{rtstt}; deployable variants use the continuous-context recipe of
\S\ref{sec:contconv}.

\noindent\textbf{Compute accounting.} MAC/frame is counted with hooks on every
convolution, linear and recurrent layer, and calibrated against hardware: the counter
reproduces the $15.08$\,M MAC/frame we independently time on the part. The profiler,
the feasibility script and the on-device reference runtime will be released.

\section{Results}
\label{sec:results}

\subsection{Deployable operating points}
Table~\ref{tab:deploy} reports the operating points. The full-band separator reaches
$4.70$\,dB cSDR at $21.9$\,M MAC/frame and runs on the part in $10.43$\,ms of the
$11.6$\,ms hop inside $1963$ of $2040$\,kB of L2 (\S\ref{sec:deploy}), so it satisfies
both constraints by measurement rather than projection. The $F{=}192$ variant trades
$0.36$\,dB for $77\%$ of compute and a wide memory margin.

\subsection{Streaming stability}
Fig.~\ref{fig:tradeoff}(b) shows what block-wise evaluation hides. Two checkpoints
differing \emph{only} in training context are run frame-by-frame with state and
convolution cache carried, as the DSP runs them. At $t{=}0$ they are
indistinguishable, because the cache genuinely is zero at stream start: the regime
chunk training creates. As real context fills it, the chunk-trained model degrades to $0.04$\,dB within ${\sim}2$\,s and stays there (a silent estimate
scores exactly $0$\,dB by construction, which is the signature), while the
continuous-trained model holds $3.17$\,dB (medians over $8$ tracks, last $30$\,s).
Block-wise, the two are indistinguishable at $3.93$\,dB. The deployed model's continuous
cSDR matches its block-wise value, and the mask holds the idle floor at exactly zero.

\subsection{Look-ahead as a latency knob}
Fig.~\ref{fig:tradeoff}(a) sweeps $Q$ on the full-band model, where the effect is
cleanest. A \emph{strictly causal} filter ($Q{=}0$) gains $0.38$\,dB cSDR and
$0.23$\,dB uSDR over the matched no-DF baseline, on \emph{every} stem: the gain is
complex FIR filtering, not future information. One look-ahead frame ($11.6$\,ms) adds
$0.19$\,dB; the aggregate then flattens ($5.50\to5.49$) while uSDR still rises
($5.35\to5.39$). Bounded look-ahead is cheap, and on the full-band model most of it is
unnecessary. The slim model qualifies that: at $Q{=}0$ it gains only $0.06$\,dB on both
metrics, reaching $+0.30$\,dB cSDR and $+0.28$\,dB uSDR only with both look-ahead frames
(Table~\ref{tab:deploy}). Cropping to $F{=}192$ leaves a strictly causal filter little to
exploit.

\noindent\textbf{What the gate learns.} Blend usage $1{-}G$ starts at $0.119$ for every
source and converges to $0.104$ (vocals), $0.021$ (other), $0.017$ (drums) and $0.007$
(bass). The near-zero bass gate is not a collapsed optimisation: overriding it at
inference on $8$ tracks moves bass in one direction only, $-0.22$\,dB forced shut and
$-0.34$, $-1.90$, $-3.80$\,dB when forced to $0.119$, $0.4$ and $0.8$. Deep filtering
does not pay on a source whose energy already sits in a few bins.

\subsection{On-device validation}
\label{sec:deploy}
The runtime is a hand-scheduled floating-point frame loop, not a generic inference
engine. The deployed full-band model runs end to end (int$\to$float, STFT, network,
high band, iSTFT, float$\to$int) in $10.43$\,ms mean and $10.44$\,ms worst case, $90\%$
of the $11.6$\,ms deadline, inside $1963$ of $2040$\,kB of L2 and $447$ of $512$\,kB of
L1. The $0.01$\,ms spread matters as much as the mean: with no allocation, cache refill
or data-dependent branch in the frame path, the deadline is met rather than usually met.
The six GRUs are $4.37$\,ms of it and the deep-filter head $2.77$. Reduced precision is
confined to where it cannot accumulate: the convolution ring caches are $16$-bit float,
halving their $1050$\,kB, but they hold a two-frame sliding window, while the weights
and the indefinitely carried GRU state stay $32$-bit. The build tracks the PyTorch
streaming twin to $7.2\times10^{-4}$ relative and runs for hours without drift. The $F{=}192$ no-DF variant is measured
too, at $7.56$\,ms ($65\%$) in $1362$\,kB, entirely $32$-bit and matching to
$1.4\times10^{-6}$; its $15.08$\,M MAC in $7.28$\,ms is the $2.07$\,$\GMACs$ used
throughout.

\section{Discussion and Limitations}
\label{sec:disc}
\noindent\textbf{We are behind on quality.} At $4.70$\,dB (bootstrap $95\%$ interval
$[4.21, 5.09]$ over the $50$ tracks) we sit $0.47$\,dB below RT-STT on our metric, or
$0.69$\,dB comparing BSSEval-v4 like for like, and are
indistinguishable from HS-TasNet. We do not claim to match RT-STT; we claim it does not
run on this class of device. Our unconstrained model reaches $5.49$\,dB but needs
$6.4\times$ the available compute.

\noindent\textbf{MAC is necessary, not sufficient.} It ignores bandwidth and
serialisation: at equal MAC a frequency-axis recurrence vectorises far worse than a
time-axis one. Pair it with measured time, as in Table~\ref{tab:deploy}.

\noindent\textbf{Scope and unmeasured quantities.} The memory verdict follows from
published parameter counts; the compute verdict for systems we did not implement rests on
the $\rho\!\approx\!1$ identity and is a lower bound. Our RT-STT MAC comes from a
reproduction scoring $4.93$\,dB, so we pair the authors' quality with our compute and
mark it. Rows $1$ and $4$ of Table~\ref{tab:deploy} are measured on hardware, the two
$F{=}192$ DF rows are not. We report no fixed-point accuracy: integer formats are not
free here, because the recurrent state \emph{is} an accumulator carried indefinitely,
which is why it stays $32$-bit on the part. We report compute, not energy, and
MUSDB18-HQ leaves domain robustness untested.

\noindent\textbf{Latency.} At $Q{=}2$ the system adds $23.2$\,ms to the $23$\,ms window:
fine for assistive listening and remixing, not for live stage monitoring, where $Q{=}0$
costs $0.24$\,dB.

\section{Conclusion}
No published real-time music source separator runs on the embedded hardware it targets,
and the reason is not model size: the two constraints eliminate different families and
parameter count predicts neither. Building one that fits was a training problem, not an
architectural one; ours runs on-device at $4.70$\,dB cSDR in $10.43$\,ms of an
$11.6$\,ms hop.

\label{endofcontent}

\bibliographystyle{IEEEbib}
\bibliography{refs}

\end{document}